\documentclass[preprint,twocolumn,5p,authoryear,compress,colorlinks=true]{elsarticle}

\journal{Icarus}

\usepackage{lineno,hyperref,siunitx,booktabs,cuted,titlesec,afterpage,xfrac}
\usepackage[version=4]{mhchem}
\modulolinenumbers[1]

\usepackage[labelfont=bf,justification=justified,singlelinecheck=false]{caption}
\makeatletter
\g@addto@macro\@floatboxreset\centering
\makeatother

\newcommand{\DOI}[1]{%
 \newline\noindent\small%
 \textit{\small{DOI:} }%
 \href{https://dx.doi.org/#1%
 }{#1%
 }%
}
\newcommand{\figref}[1]{\autoref{fig:#1}}

\begin{document}

\begin{frontmatter}
\title{The Cassini VIMS archive of Titan:\\from browse products to global infrared color maps}

\author[LPG]{St\'ephane {Le Mou\'elic}\corref{correspondingauthor}}\ead{stephane.lemouelic@univ-nantes.fr }
\author[ESAC]{Thomas Cornet}
\author[IPGP]{S\'ebastien Rodriguez}
\author[JPL]{Christophe Sotin}
\author[LPG]{Beno\^{i}t Seignovert}
\author[DPUI]{~\\Jason W. Barnes}
\author[DPS]{Robert H. Brown}
\author[JPL]{Kevin H. Baines}
\author[JPL]{Bonnie J. Buratti}
\author[PSI]{Roger N. Clark}
\author[Cornell]{~\\Philip D. Nicholson}
\author[IRAP]{J\'er\'emie Lasue}
\author[DPS]{Virginia Pasek}
\author[MIT]{Jason M. Soberblom}

\address[LPG]{LPG, UMR 6112, CNRS, Universit\'e de Nantes, 2 rue de la Houssini\`ere, 44322 Nantes, France}
\address[ESAC]{European Space Astronomy Centre (ESA/ESAC), Villanueva de la Canada, Madrid, Spain}
\address[IPGP]{IPGP, CNRS-UMR 7154, Universit\'e Paris-Diderot, USPC, Paris, France}
\address[JPL]{Jet Propulsion Laboratory, California Institute of Technology, Pasadena, CA 91109, USA}
\address[DPUI]{Department of Physics, University of Idaho, Engineering-Physics Building, Moscow, ID 83844, USA}
\address[DPS]{Department of Planetary Sciences, University of Arizona, Tucson, AZ 85721, USA}
\address[PSI]{Planetary Science Institute, Tucson, USA}
\address[Cornell]{Department of Astronomy, Cornell University, Ithaca, NY 14853, USA}
\address[IRAP]{IRAP, Toulouse, France}
\address[MIT]{MIT, Department of Earth, Atmospheric and Planetary Sciences, Cambridge, MA 02139, USA}

\cortext[correspondingauthor]{Corresponding author}

\begin{abstract}
We have analyzed the complete Visual and Infrared Mapping Spectrometer (VIMS) data archive of Titan.
Our objective is to build global surface cartographic products, by combining all the data gathered during the 127 targeted flybys of Titan into synthetic global maps interpolated on a grid at \SI{32}{pixels} per degree (\SI{\sim 1.4}{km/pixel} at the equator), in seven infrared spectral atmospheric windows. Multispectral summary images have been computed for each single VIMS cube in order to rapidly identify their scientific content and assess their quality.
These summary images are made available to the community on a public website (\href{https://vims.univ-nantes.fr/}{vims.univ-nantes.fr}).
The global mapping work faced several challenges due to the strong absorbing and scattering effects of the atmosphere coupled to the changing observing conditions linked to the orbital tour of the Cassini mission. We determined a surface photometric function which accounts for variations in incidence, emergence and phase angles, and which is able to mitigate brightness variations linked to the viewing geometry of the flybys.
The atmospheric contribution has been reduced using the subtraction of the methane absorption band wings, considered as proxies for atmospheric haze scattering. We present a new global three color composite map of band ratios (red: \SI{1.59/1.27}{\um}; green: \SI{2.03/1.27}{\um}; blue: \SI{1.27/1.08}{\um}), which has also been empirically corrected from an airmass (the solar photon path length through the atmosphere) dependence. This map provides a detailed global color view of Titan's surface partially corrected from the atmosphere and gives a global insight of the spectral variability, with the equatorial dunes fields appearing in brownish tones, and several occurrences of bluish tones localized in areas such as Sinlap, Menvra and Selk craters. This kind of spectral map can serve as a basis for further regional studies and comparisons with radiative transfer outputs, such as surface albedos, and other additional data sets acquired by the Cassini Radar (RADAR) and Imaging Science Subsystem (ISS) instruments.
\end{abstract}

\begin{keyword}
Titan \sep Titan surface \sep Image processing \sep Infrared observations
\DOI{10.1016/j.icarus.2018.09.017}
\end{keyword}

\end{frontmatter}


\section{Introduction}

\begin{figure*}[!ht]
 \includegraphics[width=\linewidth]{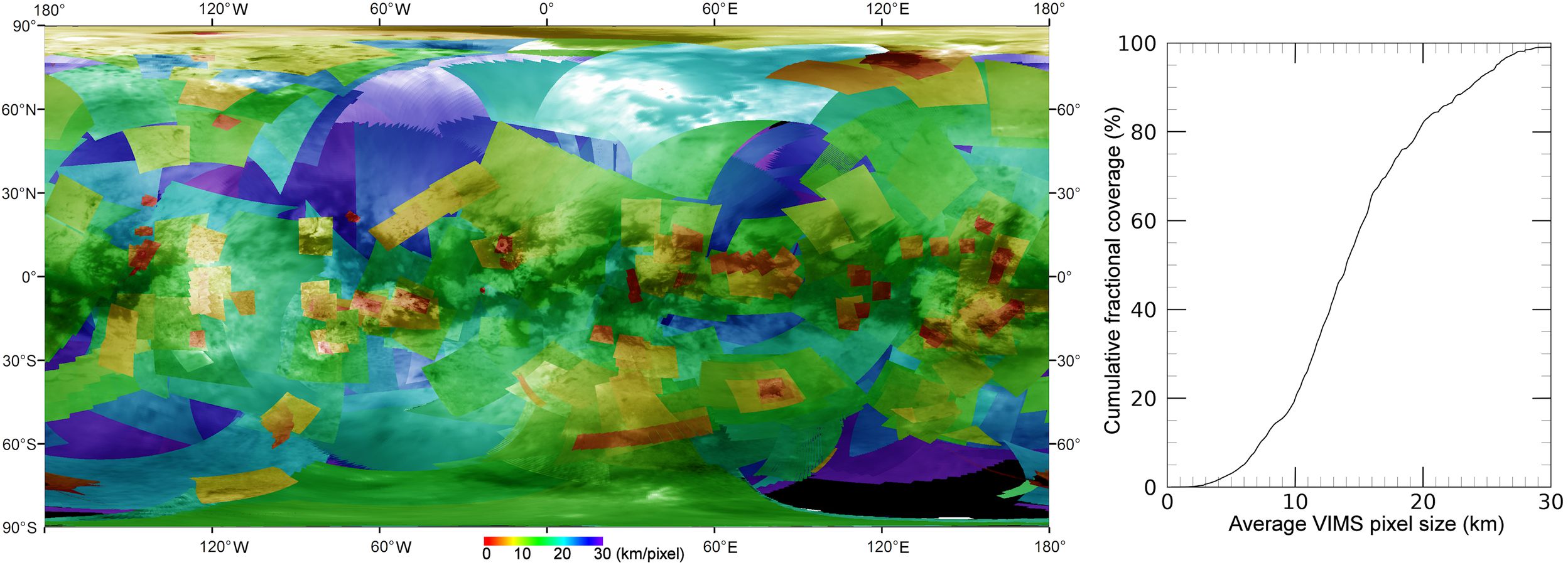}
 \caption{Pixel size (in km) of the cubes used to build our global mosaic. A global map at \SI{2}{\um} is shown in transparency. The cumulative fractional coverage is shown on the right. Only \SI{5}{\percent} of the surface has been observed with a pixel size better than \SI{6}{km}. \SI{60}{\percent} was observed in conditions better than \SI{15}{km/pixel}.}
 \label{fig:fig_1}
\end{figure*}

Titan has been recognized since the era of the Voyager space missions as one of the most interesting bodies in the field of comparative planetology. Although the surface is totally masked in the visible wavelength by scattering and absorptions in the atmosphere, the geological diversity of Titan has been progressively revealed by the instruments onboard the Cassini spacecraft, which spent 13 years between 2004 and 2017 in the Saturnian system. The Radar instrument onboard Cassini was able to observe directly through Titan's atmosphere using a centimetric wavelength \citep{Elachi2005}. Optical observations were performed by two other instruments. The Imaging Science Subsystem (ISS), composed of two multispectral framing cameras, provided information on the surface thanks to its 0.\SI{93}{\um} CB3 filter \citep{Porco2005}. The Visual and Infrared Mapping Spectrometer (VIMS) acquired hyperspectral images which gave access to the surface through partially transparent atmospheric windows in the infrared at \num{1.08}, \num{1.27}, \num{1.59}, \num{2.01}, \num{2.69}, \num{2.78} and \SI{5}{\um} \citep{Brown2004,Sotin2005}. The spectral dimension of VIMS provides the possibility to retrieve information on compositional and/or physical state (grain size) variations at the surface, in addition to giving access to clouds and aerosols properties. With observations collected year after year, the geological diversity of Titan proved to be exceeding the most optimistic expectations. Earth-like processes and landforms such as cloud formation, river flowing at the surface (implying rainfalls), polar lakes and seas, mountain chains, equatorial dunes fields, impact craters, were progressively discovered and characterized \citep{Tomasko2005,Stofan2007,Radebaugh2007,Radebaugh2008,Wood2010,Aharonson2014}. The main difference with Earth comes from the nature of the materials: with an average surface temperature of \SI{-180}{\degreeCelsius}, methane is close to its triple point, playing on Titan the role of water on Earth. Only very few impact craters have been observed on the entire surface \citep{Wood2010}, which indicates that the surface is geologically relatively young, probably reprocessed by tectonic events, erosion of the bedrock, and deposition of sediments from air fall or slope/fluvial transport processes \citep{Neish2016,Brossier2018}.

In this paper, we focus our study on the VIMS global archive, with the objective of producing global color mosaics of the complete data set of Titan acquired between T0 (July 2004) and the last targeted flyby, T126, in April 2017. The correspondence between the Cassini flybys of Titan and the Cassini orbits around Saturn can be found in \cite{Seignovert2015}. We consolidate a previous study, which was limited to data acquired up to June 2010 only \citep{LeMouelic2012a}. Merging data acquired in very different viewing conditions into global homogeneous maps is a challenge due to the presence of the atmosphere, which induces strong absorbing and scattering effects when coupled with the changing geometry of the flybys. In a first section, we describe the global VIMS data set, its observing modes and radiometric calibration, and discuss the production of multispectral summary images designed to catch the scientific content of each observation in an optimized way. In a second step, we present how the data were merged into global maps, after implementing empirical corrections for the surface photometry and for the atmospheric effects. We discuss in particular the use of band ratios, before concluding with series of orthographic views and a focus on the Huygens landing site.

\section{Description of the VIMS dataset}

\begin{figure*}[!ht]
 \includegraphics[width=\linewidth]{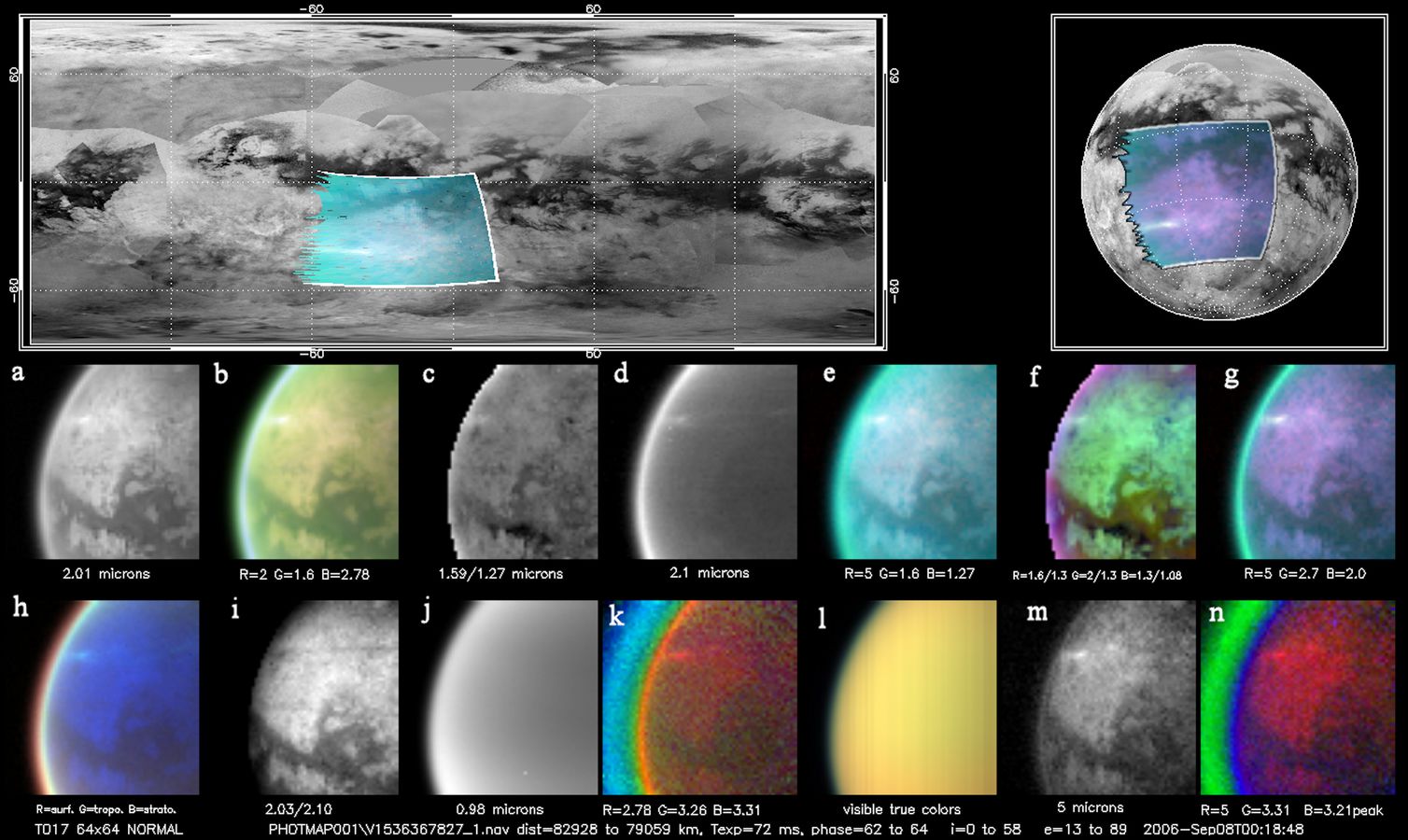}
 \caption{Example of VIMS multispectral summary image on cube CM\_1536367827 (08 September 2006). Black and white images and color composites have been designed to catch variations linked both to surface and atmospheric features.}
 \label{fig:fig_2}
\end{figure*}

\subsection{VIMS observing modes}

The Visual and Infrared Mapping Spectrometer (VIMS) onboard Cassini acquired up to \SI{64x64}{pixels} images in 352 spectral channels from \num{0.35} to \SI{5.12}{\um} \citep{Brown2004}.
VIMS was composed of two separate instruments.
The first was a two-dimension CCD array that covers the visible range (\SIrange{0.35}{1.04}{\um}) with 96 spectral channels.
The second covered the infrared range (\SIrange{0.88}{5.12}{\um}) with 256 channels on a linear detector array and a bidirectional mirror (whisk-broom).
The visible part has proved to be very challenging to observe the surface of Titan, due to the strongly absorbing and scattering atmosphere \citep{Tomasko2005,Hirtzig2009,Vixie2012}.
Since we focus our present study on surface observations, we therefore center our efforts on the infrared detector. Between 2004 and 2017, hyperspectral data have been gathered during 127 targeted Titan close encounters, in addition to more distant untargeted observations.
The spatial size of data cubes were optimized to take advantage of any acquisition opportunities, which relied on which instrument was driving the pointing of Cassini during the closest approach phase.
Occasionally, cubes were acquired on a line mode ({\emph{noodle}), letting the spacecraft drift build the second dimension of the image after concatenation of a series of hundreds of line cubes.
An occultation mode was also designed to catch the signal of stars crossing Titan's atmosphere.
The time exposure generally ranged from \SI{13}{ms} (used mostly at closest approach to compensate the fast drift of the surface) to \SI{640}{ms} (when a higher S/N ratio was desired to look for subtle spectral signatures at long wavelengths).
More than \num{60000} hyperspectral cubes of Titan have been acquired during the entire Cassini mission, with a pixel size as fine as \SI{500}{meters} at best when VIMS was operating right at closest approach in very few occasions. \figref{fig_1} shows the spatial coverage obtained with all cubes acquired within thresholds that we describe in a later stage to build the mosaics, with pixel sizes smaller than \SI{30}{km} and with time exposures in the \SIrange{20}{300}{ms} range to avoid low signal to noise ratios and saturated cubes. The panel on the right displays the corresponding cumulative fractional coverage. The region around (\ang{80}S, \ang{120}E), which represents \SI{\sim 1}{\percent} of the surface, was never observed within these thresholds. Only \SI{5}{\percent} of the surface was covered with a pixel scale lower than \SI{6}{km/pixel}. The cumulative global coverage raises to \SI{20}{\percent} when considering observations better than the \SI{10}{km/pixel} scale, and \SI{60}{\percent} for observations better than the \SI{15}{km/pixel} scale.

\subsection{Radiometric calibration}
All the data cubes have been calibrated in reflectance factor $I/F$, where F is the solar flux (irradiance) and $I$ is the calibrated radiance measured by VIMS. We followed the VIMS pipeline described in \cite{Brown2004} and \cite{Barnes2007}, and further refined using a time-dependent radiometric calibration aimed at correcting a small wavelength shift that has been identified during the last years of the mission \citep{Clark2018}. Up to \SI{\sim 10}{nm} of progressive shift is observed when comparing data taken in 2004 and data taken in 2017 \citep{Clark2018}. Despite this shift being small, it can dramatically alter the surface information in the sharp atmospheric windows (especially at short wavelength), and produce significant seams if left uncorrected. In the last calibration step, all spectra of the mosaic have therefore been converted to a common reference wavelength of 2004 with a spline interpolation, using the shifts evaluated by \citep{Clark2018}.

\subsection{VIMS multispectral summary products}
For each VIMS hyperspectral cube of Titan (except the single line cubes and the cubes taken in occultation mode), we have setup a multispectral summary image designed to highlight the spectral diversity of the observation using specific combination of channels, and displaying the cubes under different map projections.
This strategy is similar to the one used by the Compact Reconnaissance Imaging Spectrometer for Mars (CRISM) team to automatically search for mineral signatures on Mars \citep{Pelkey2007}.
It allows us to visually find the most interesting data, in addition to identify corrupted cubes. \figref{fig_2} shows an example on a typical multispectral summary product acquired at T17 in September 2006. We choose to display both enhanced black and white and color composites which are dedicated either to surface or atmospheric features. We use for example the \SI{2.01}{\um} channel (\figref{fig_2}a) to emphasize the details of the surface.
We also calculated three RGB composites of single bands inspired from previous studies \citep{Barnes2007, Soderblom2009a, LeMouelic2012a}, which reveal at the same time clouds and surface features (\figref{fig_2}b/e/g).
The \SI{1.59/1.27}{\um} ratio (\figref{fig_2}c) emphasizes spectral variations of the surface. The image acquired at \SI{2.1}{\um} (\figref{fig_2}d) is used to detect clouds \citep{Rodriguez2009, Rodriguez2011, Turtle2018}, which appear bright at this wavelength where the surface is not seen.
\figref{fig_2}f corresponds to an RGB color composite of the \SI{1.59/1.27}{\um}, \SI{2.03/1.27}{\um} and \SI{1.27/1.08}{\um} band ratios respectively, which provide the most sensitivity to surface heterogeneities.
We included in \figref{fig_2}h a color composite with the surface, tropospheric and stratospheric parameters of \cite{Brown2010}.
The \SI{2.03/2.10}{\um} ratio (\figref{fig_2}i) provides a tentative normalization of the illuminating conditions.
The image acquired at \SI{0.98}{\um} (\figref{fig_2}j) shows a pure atmospheric scattering observation. We have added two color composites (R=\SI{2.78}{\um}, G=\SI{3.26}{\um}, B=\SI{3.31}{\um} in \figref{fig_2}k and R=\SI{5}{\um}, G=\SI{3.31}{\um}, B=\SI{3.21}{\um} in \figref{fig_2}n) which are sensitive to the methane fluorescence and reveal the layers of the atmosphere.
A true color image is also displayed (\figref{fig_2}l).
Finally, the image acquired at \SI{5}{\um} (\figref{fig_2}m) corresponds to the average of all channels between \SI{4.90} and \SI{5.12}{\um}.
This wavelength range is the least affected by atmospheric scattering. Information regarding the flyby number, date, distance range of Cassini to Titan's surface at the time of the cube acquisition, pixel exposure time, ranges for the phase, incidence and emergence angles is also displayed.

The browse products contain a wealth of information that could potentially stimulate further focused atmospheric and surface studies. A dedicated website has been setup to provide a user-friendly access to all these multispectral summary images, produced from the Planetary Data System archive and covering the entire mission (\href{https://vims.univ-nantes.fr/}{vims.univ-nantes.fr}).

\section{Merging data into global maps}

\begin{figure*}[!ht]
 \includegraphics[width=\linewidth]{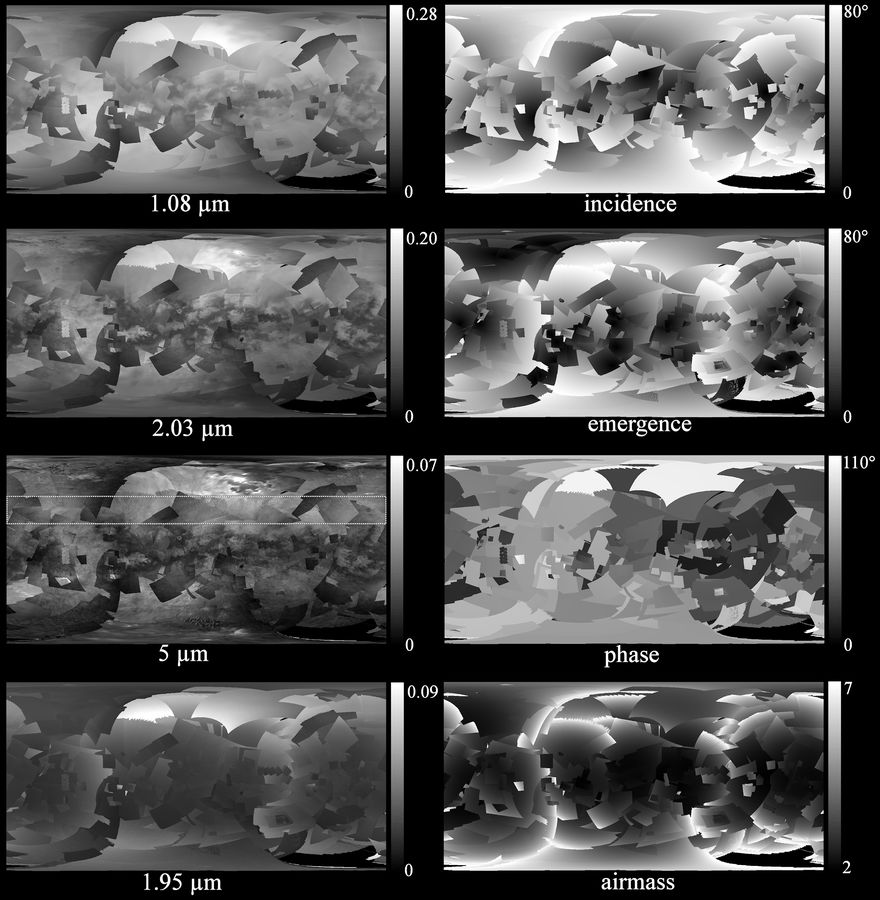}
 \caption{Raw global mosaics in equidistant cylindrical projection at \num{1.08}, \num{2.03}, \SI{5}{\um} (surface windows) and \SI{1.95}{\um} (atmosphere only) compared to global mosaics of incidence, emergence, phase, and airmass, containing all the VIMS cubes acquired during the entire Cassini mission, within the data filters described in the text. The observing conditions vary widely during the mission, which causes significant seams or boundaries to appear in the uncorrected mosaics. The white dashed rectangle on the \SI{5}{\um} map corresponds to the test area used to determine the corrections.}
 \label{fig:fig_3}
\end{figure*}

Our final objective is to produce VIMS synthetic global maps interpolated on a grid at \SI{32}{pixels} per degree (corresponding to a spatial sampling of \SI{\sim 1.4}{km} at the equator), using different combinations of wavelengths emphasizing surface spectral heterogeneities. Merging 13 years of data represents a significant challenge considering the non-negligible contribution of Titan's absorbing and scattering atmosphere even in the methane optical windows, the huge variations in observing conditions encountered throughout the mission, as well as the temporal changes that could have occurred in the surface and atmosphere during 13 years \citep{Barnes2013b, Solomonidou2016}.

\subsection{Data fusion strategy}
In order to build global maps, individual data cubes have been sorted by increasing spatial resolution, with the high resolution images on top of the mosaic and the low resolution images used as background.
Other strategies could be envisaged to give more weight to other parameters influencing data quality, such as time exposure or low airmass (that we defined by $\sfrac{1}{\cos i} + \sfrac{1}{\cos e}$ in the plane-parallel atmosphere approximation), instead of the spatial resolution only \citep[\emph{i.e.,}][]{Barnes2007}.
After testing this approach, we decided to keep the spatial resolution as the main criterion to emphasize the finest details of the surface. We filtered out the observing geometry in order to remove the pixels acquired in too extreme illuminating and viewing conditions, which produce strong seams in the VIMS mosaics due to enhanced surface and atmospheric photometric effects. We used thresholds of \ang{80} both on the incidence and emergence angles, \ang{110} on the phase angle, and 7 on the airmass. These thresholds correspond to a trade-off between the surface coverage (in particular in polar areas, most often viewed at extreme geometries) and the mosaic quality. The lowest values of the incidence, emergence, phase and airmass in the mosaic are \ang{0.12}, \ang{0.02}, \ang{11.1}, \num{2.01} respectively. The exposure time has been restrained to the \SIrange{20}{300}{ms} range in order to avoid cubes with low signal-to-noise ratio and saturated data.

\figref{fig_3} shows the resulting global mosaic of relevant geometric viewing parameters (incidence, emergence, phase and airmass), the $I/F$ at \num{1.08}, \num{2.03} and \SI{5}{\um} (surface windows), the $I/F$ at \SI{1.95}{\um} (where the atmosphere is not transparent) with no correction for geometry nor for the atmospheric effects.
Many boundaries or seams appear between individual images in these raw mosaics. They are mainly caused by the varying viewing angles (incidence, emergence, phase) between data acquired during the different flybys, which induce strong atmospheric and surface photometric effects. A dependence with airmass is also observed. Other discrepancies might exist due to surface and atmospheric temporal variations, and residual calibrations artifacts.

\subsection{Photometric correction at 5 microns}\label{ssec:photometric_correction}

In order to account for the variations of solar illumination (incidence $i$) and viewing angle (emergence $e$) between different flybys, a surface photometric correction had to be implemented. We focused on the \SI{5}{\um} atmospheric window, which is the least affected by atmospheric scattering and absorption, and thus is the most sensitive to surface photometric effects. We selected a test area presenting a rather homogeneous brightness at \SI{5}{\um}, located in the Northern mid-latitudes between \ang{37.5}N and \ang{52.5}N (white dashed rectangle in the \SI{5}{\um} map of \figref{fig_3}), outside of the lakes, seas, or dune fields. Given the relatively weak atmospheric absorption and haze scattering in this window, most of the variations seen in this portion of the mosaic come from the viewing conditions of the surface.

We have tested several surface photometric corrections commonly found in the literature. These include the Lambert, Lommel-Seeliger and Lunar Lambert disc functions, coupled with different particle phase functions $P(\phi)$ such as the Rayleigh, Henyey-Greenstein and Hapke lunar functions \cite{Hapke2012}. While particle phase functions are usually used to infer particle shape and size properties (\emph{e.g.,} Henyey-Greenstein two lobes phase function), we rather focus on the mosaic enhancement, the phase function giving us an indication of the degree of anisotropy in scattering.
Whereas a pure Lambert function gave satisfactory results to perform a first order correction of the incidence angle ($i$) in an earlier version of the maps \citep{LeMouelic2012a}, we realized that a supplementary correction of the emergence ($e$) and phase ($\phi$) angles was needed to account for the extreme diversity of the viewing conditions, in particular on the northern polar regions, which were only seen during the second half of the Cassini mission after the dissipation of the northern cloud and haze \citep{LeMouelic2012b, LeMouelic2018}.
These areas were not included in our previous maps. Our best result was obtained with a Lunar Lambert type function (Eqs. \eqref{eq:1} and \eqref{eq:2}) with a lunar-like weighting factor $A = 0.285$ (\figref{fig_4} and \ref{fig:fig_5}).

\begin{align}
\label{eq:1}
    f = A \cdot \frac{\cos i}{\cos i + \cos e} \cdot P(\phi) + \left(1 - A \right) \cdot \cos i \\
\label{eq:2}
    P(\phi) = \frac{4\, \pi}{5} \left[ 
        \frac{\sin\phi + (\pi - \phi) \cos\phi}{\pi} +
        \frac{\left(1 - \cos\phi\right)^2}{10}
    \right]
\end{align}

$P(\phi)$ in equation \eqref{eq:1} is the single-particle phase function of the surface. The lunar theoretical particle phase function of \cite{Hapke1963} provided satisfactory results to describe this term (equation \eqref{eq:2}), as it was already noted by \cite{Cornet2012}. The fact that the point cloud in \figref{fig_4} is aligned with the origin of the graph confirms the hypothesis that the additive scattering term at \SI{5}{\um} is negligible.

\begin{figure}[!ht]
 \includegraphics[width=\linewidth]{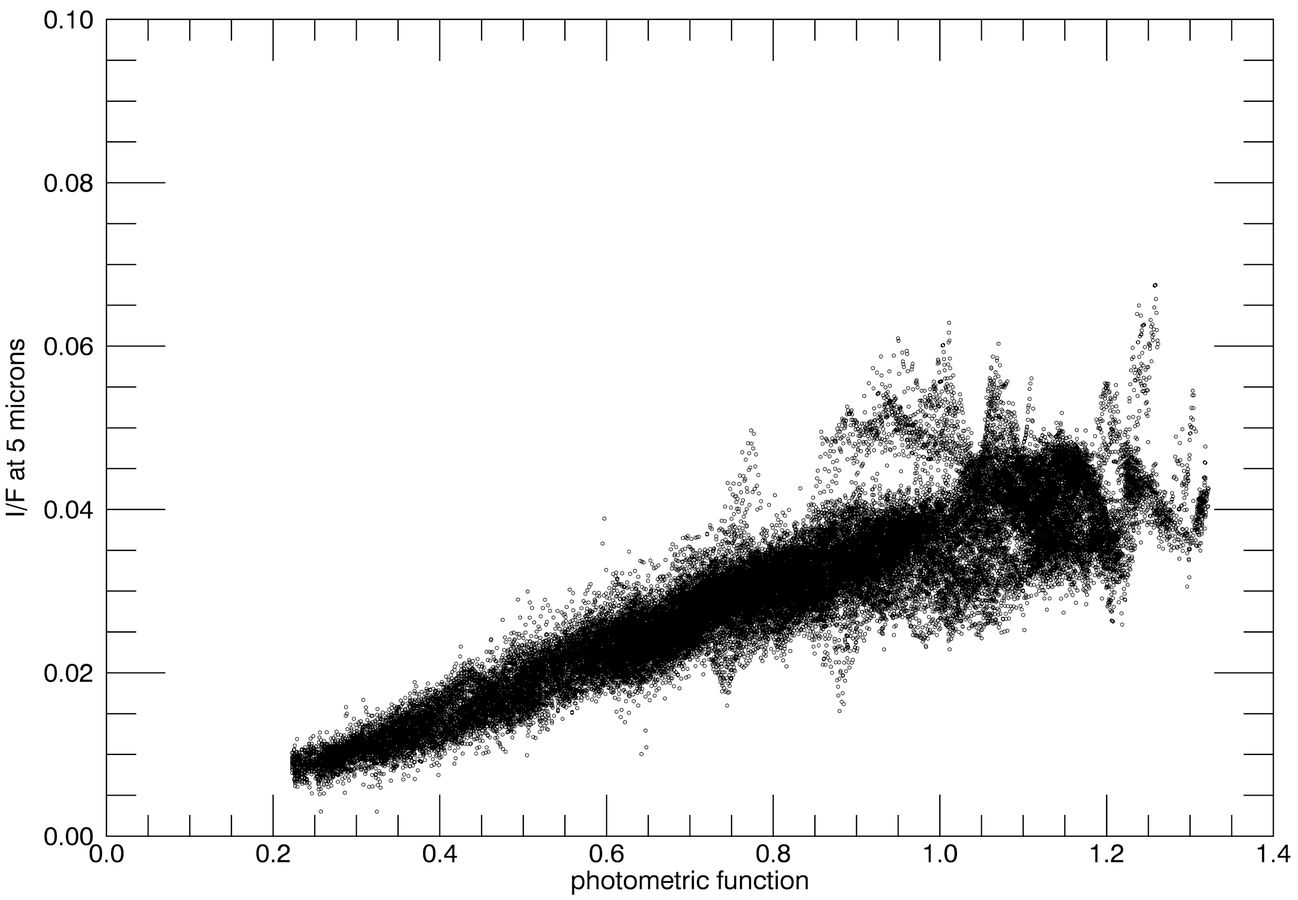}
 \caption{$I/F$ at \SI{5}{\um} versus the photometric function described in equation \eqref{eq:1}. The linear correlation shows that this function can be used to correct at first order from the effect of incidence, emergence and phase.}
 \label{fig:fig_4}
\end{figure}

\begin{figure}[!ht]
 \includegraphics[width=\linewidth]{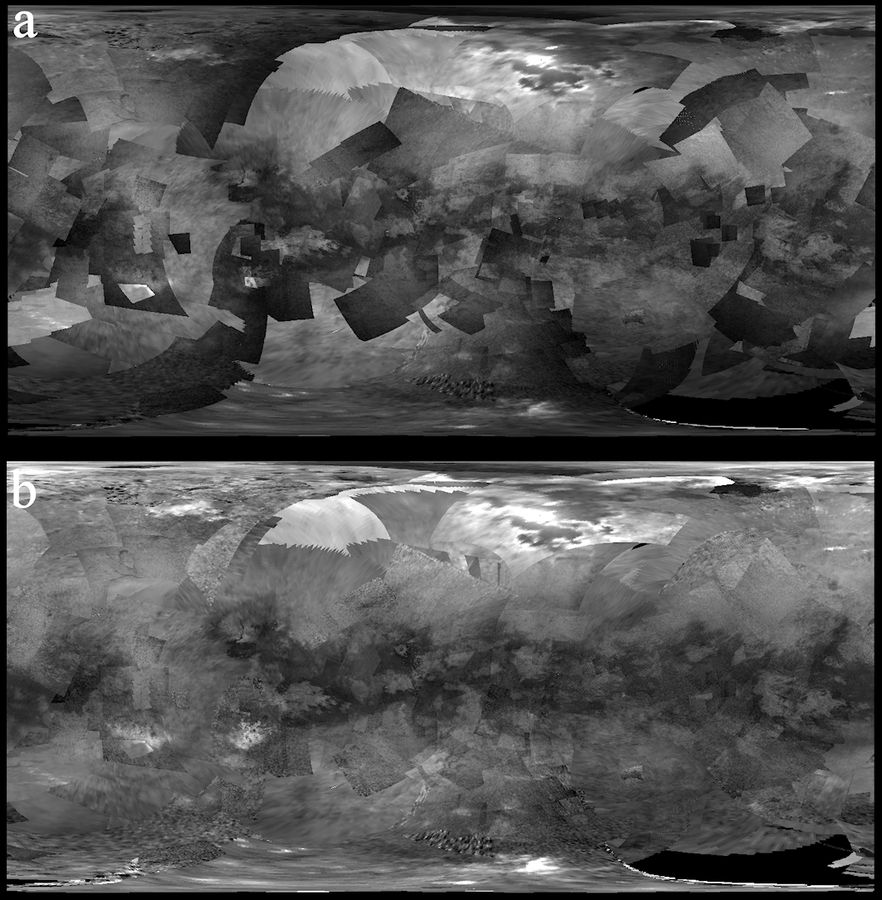}
 \caption{(a) Uncorrected global map at \SI{5}{\um}. (b) \SI{5}{\um} map corrected using the surface photometric function described in equations \eqref{eq:1} and \eqref{eq:2}. Most of the seams have been smoothed out, except on two cubes in northern regions, one of which exhibiting a broad specular reflexion on Kraken Mare.}
 \label{fig:fig_5}
\end{figure}

The uncorrected map at \SI{5}{\um} is shown in \figref{fig_5}a. The map at \SI{5}{\um} corrected for the photometry with the factor described in equations \eqref{eq:1} and \eqref{eq:2} is shown in \figref{fig_5}b. We see that the level of seams has significantly decreased in almost all regions, except two cubes in northern latitudes taken in an extreme geometry and which contain in particular a broad specular reflexion on Kraken Mare. Very bright features in this map correspond to possible evaporites \citep{Barnes2011, MacKenzie2014}, specular reflections on the northern seas \citep{Sotin2012, Soderblom2012, Barnes2013a, Barnes2014}, possible cryovolcanic candidates \citep{Lopes2013}, or unfiltered clouds \citep{Turtle2018}.

In the following, we will use the same photometric function for all other surface windows, assuming that the surface properties are not wavelength-dependent. This is a shortcoming, as a complete solution would require to derive different photometric parameters for each wavelength, as it is commonly done for example on bodies such as Mars \citep{Binder1972} and the Moon \citep{Lane1973}. However, to achieve this, the contribution of the atmosphere on Titan has to be fully removed prior to the photometric parameters computation, which still makes it very challenging at this stage, and falls beyond the scope of this paper. We leave this issue to further studies based on complete radiative transfer approaches.

\subsection{Short wavelengths case}

\begin{figure*}[!ht]
 \includegraphics[width=.99\linewidth]{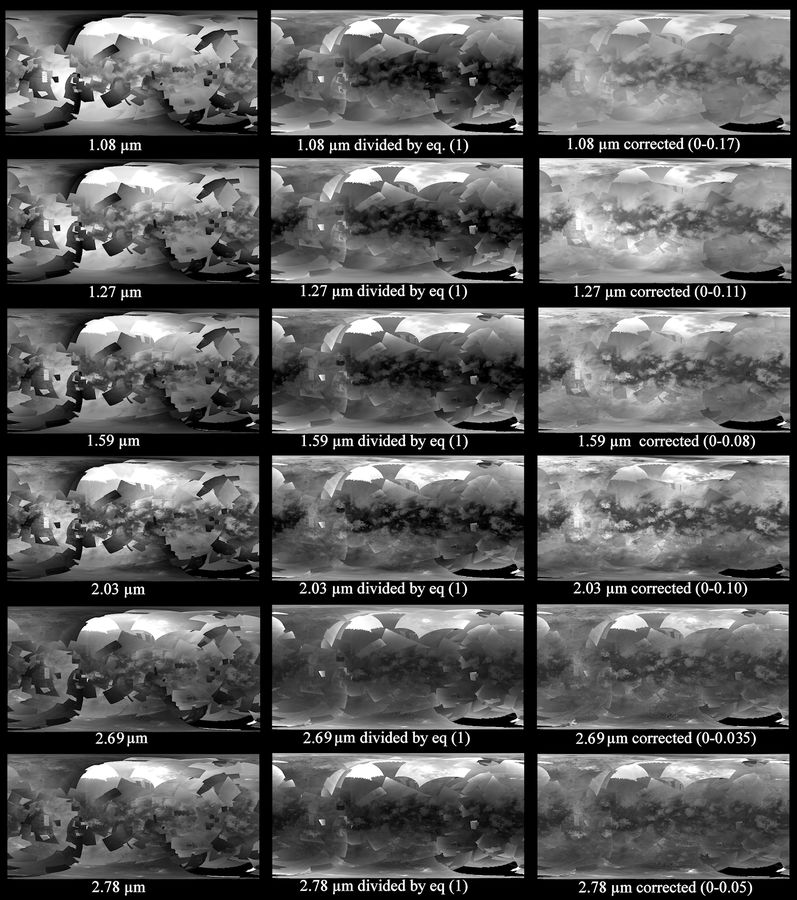}
 \caption{Mosaics at \num{1.08}, \num{1.27}, \num{1.59}, \num{2.03}, \num{2.69}, \SI{2.78}{\um} without correction (left), with a photometric correction only (center), and with the first order correction of the additive scattering term prior to the photometric correction (right).}
 \label{fig:fig_6}
\end{figure*}

Whereas the \SI{5}{\um} methane window is almost free of atmospheric scattering, this is not the case for wavelengths shorter than 3 microns, which contain an additive scattering contribution of the aerosols. To mitigate this effect, we use the wings of the atmospheric windows as a proxy to correct for the amount of additive scattering present in the center of these windows, where the surface is seen by VIMS. This process is already described in \cite{LeMouelic2012a} and will therefore not be fully reproduced here. We used the same set of k-factor (values of \num{1.15}, \num{1.50}, \num{1.60}, \num{1.29} and \num{1.14} respectively for the \num{1.08}, \num{1.27}, \num{1.59}, \num{2.03} and \SI{2.78}{\um} atmospheric windows, \emph{e.g.} Tab. 1 in \cite{LeMouelic2012a}, which account for the difference of transparency between the center of the windows and their wings. The wings were taken at \num{1.03}, \num{1.14}, \num{1.22}, \num{1.32}, \num{1.49}, \num{1.65}, \num{1.95}, \num{2.13}, \num{2.64} and \SI{2.83}{\um}.
These wavelengths correspond the first images (departing from the center of the windows) for which no surface feature is visually detectible in the global mosaic, even at the lowest airmass conditions. The \SI{2.69}{\um} image is corrected with the same wings and k-factor as the \SI{2.78}{\um} image, as these two windows correspond to a double peak rather than a single narrow one.

\figref{fig_6} presents a comparison of the maps in the surface windows before (left column) and after this empirical atmospheric correction process (right column). The middle column shows partial results, where only the surface photometric correction described in section \ref{ssec:photometric_correction} has been applied without the subtraction of the band wings, which would be the typical correction for data acquired on an airless body. The level of residual seams has been decreased in all the windows in most cases after the complete correction process (right column).

\begin{figure}[!ht]
 \includegraphics[width=\linewidth]{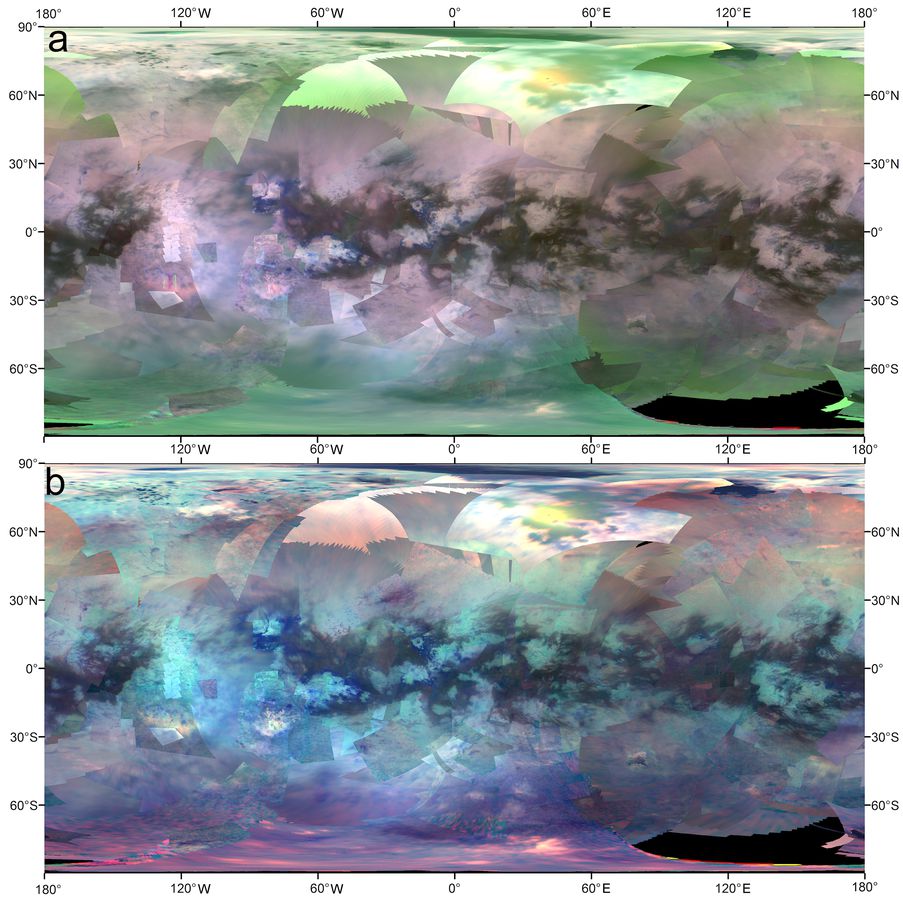}
 \caption{(a) RGB global map with the red, green and blue controlled by the \num{2.0}, \num{1.59} and \SI{1.27}{\um} channels respectively. (b) RGB global map with the red, green and blue controlled by the \num{5}, \num{2.0} and \SI{1.27}{\um} channels respectively.}
 \label{fig:fig_7}
\end{figure}

In order to emphasize spectral variations linked to compositional heterogeneities, the maps in \figref{fig_6} can be combined into RGB color composites. \figref{fig_7} shows two RGB global color maps. \figref{fig_7}a has been coded with the red, green and blue channels controlled respectively by the \num{2.01}, \num{1.59} and \SI{1.27}{\um} mosaics empirically corrected from atmospheric scattering and photometry with the method described above. \figref{fig_7}b corresponds to a global mosaic with the red, green and blue channels controlled by the \num{5}, \num{2.01} and \SI{1.27}{\um} images respectively. Introducing the \SI{5}{\um} window decreases the level of the greenish trend seen in \figref{fig_7}a due to the residual absorption and scattering in the atmosphere, mainly in the polar regions. A broad specular reflexion is seen on Kraken Mare in both cases.

These color composites have been widely used in regional studies \citep[\emph{e.g.,}][]{Barnes2007, Barnes2011, Soderblom2009b, Soderblom2009a, Cornet2012, Rodriguez2014}. Our objective is now to go one step further by investigating band ratios, which are extremely sensitive to subtle spectral variations, as it has already been shown in the case of airless bodies.

\subsection{Color composites of band ratios}

To better emphasize spectral heterogeneities, we also computed RGB composites of band ratios. Band ratios, which cancel out all multiplicative factors in absence of additive components, is a powerful technique widely used in planetary sciences. For Titan, the \SI{1.59/1.27}{\um}, \SI{2.03/1.27}{\um} and \SI{1.27/1.08}{\um} ratios proved to be useful for localized regional studies \citep{LeMouelic2008, Brossier2018}. However, using band ratios on global maps of Titan still remains very challenging, as ratios are generally much more sensitive to atmospheric effects and any residual calibration artifacts than RGB composites of single bands only. Producing fully artifact-free global maps of band ratios would still be some sort of ultimate cartographic product that requires a thorough investigation of the residuals present in the corrected mosaics. 

\begin{figure}[!ht]
 \includegraphics[width=\linewidth]{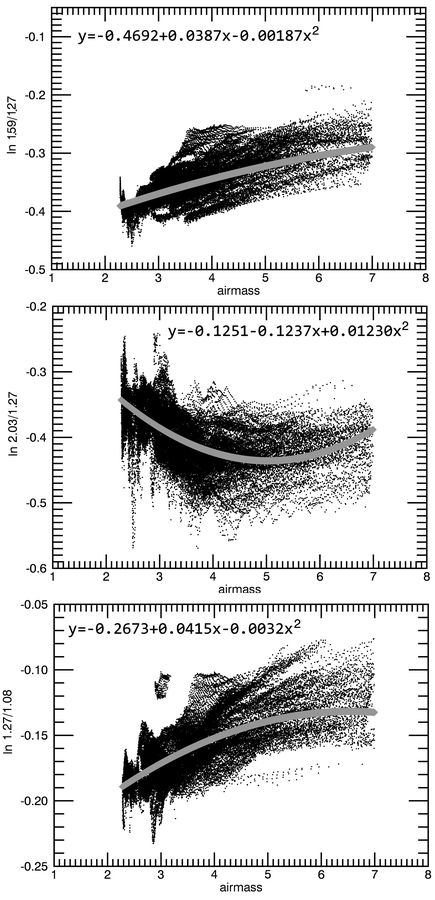}
 \caption{Dependence of band ratios with airmass on the test area containing mostly homogeneously bright terrains located in the latitudinal belt between \ang{37.5}N and \ang{52.5}N. A positive correlation is observed for the upper and lower panels, whereas the middle panel presents a negative correlation at low airmass values.  This dependence can be empirically corrected at first order using a second order polynomial fit.}
 \label{fig:fig_8}
\end{figure}

In order to make progress in this direction, we investigated the dependence of the ratios with geometric parameters. We observed in particular that the logarithm of each ratio appears correlated with our airmass parameter, so with the amount of atmosphere that the light has been crossing. This is illustrated in the scatter plots of \figref{fig_8}, corresponding to all the points located in our test latitudinal belt between \ang{37.5}N and \ang{52.5}N, containing mostly homogeneously bright terrains. The dependence of the ratios with the airmass is also apparent when comparing the RGB composite of the \SI{1.59/1.27}{\um}, \SI{2.03/1.27}{\um} and \SI{1.27/1.08}{\um} ratios in \figref{fig_9}a and the airmass map in \figref{fig_3}. This is particularly the case in fuzzy pinkish areas seen near the equator in \figref{fig_9}a.

\begin{figure*}[!ht]
 \includegraphics[width=\linewidth]{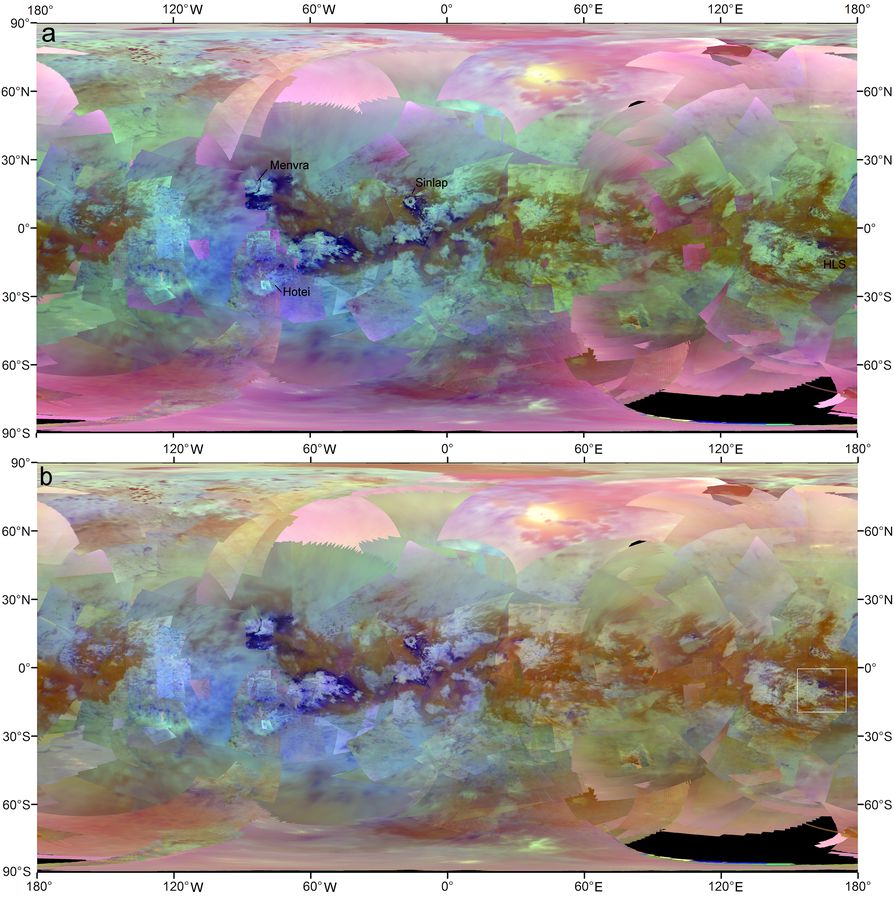}
 \caption{(a) RGB color map with the red, green and blue channels controlled by the \SI{1.59/1.27}{\um}, \SI{2.03/1.27}{\um} and \SI{1.27/1.08}{\um} ratios respectively. (b) Same mosaic after the empirical correction of the airmass dependence (equations \eqref{eq:3}, \eqref{eq:4} and \eqref{eq:5}). The correction decreases the contrast at the seams due to the atmosphere at the equator and near the poles. The white rectangle shows the location of the zoom on the Huygens landing site displayed in \figref{fig_11}}
 \label{fig:fig_9}
\end{figure*}

Following the systematic trends observed in \figref{fig_8}, we decided to empirically remove this dependence with airmass using a second order polynomial fit derived on the scatter plot of the logarithm of the ratios versus the airmass. The corresponding correction formula are given below:

\begin{align}
\label{eq:3}
    \left( \frac{R_{1.59}}{R_{1.27}} \right)_\textrm{corrected} &= 
    \left( \frac{R_{1.59}}{R_{1.27}} \right) \exp^{
        -\left( 0.0387 a - 0.00187 a^2\right)
    }
\\
\label{eq:4}
    \left( \frac{R_{2.03}}{R_{1.27}} \right)_\textrm{corrected} &= 
    \left( \frac{R_{2.03}}{R_{1.27}} \right) \exp^{
        -\left( -0.1237 a - 0.0123 a^2\right)
    }
\\
\label{eq:5}
    \left( \frac{R_{1.27}}{R_{1.08}} \right)_\textrm{corrected} &= 
    \left( \frac{R_{1.27}}{R_{1.08}} \right) \exp^{
        -\left( 0.0415 a - 0.0032 a^2\right)
    }
\end{align}

where $R_\lambda$ is the reflectance at the wavelength $\lambda$ and $a$ is the airmass defined by $\sfrac{1}{\cos i} + \sfrac{1}{\cos e}$. After this purely empirical step, the resulting RGB color composite of band ratios appears much less dependent on the geometry of observations, as shown in \figref{fig_9}b. Whereas this map still contains some dependence with atmospheric contributions and pure brightness variations, subtle color differences are strongly emphasized compared to previous RGB maps due to the use of the ratios. The strongly diffusing atmosphere still hampers the study of the polar areas (appearing in pink), but the ratios nicely reveal the extent of equatorial dune fields, which appear in brownish tones.

\begin{figure*}[!ht]
 \includegraphics[width=\linewidth]{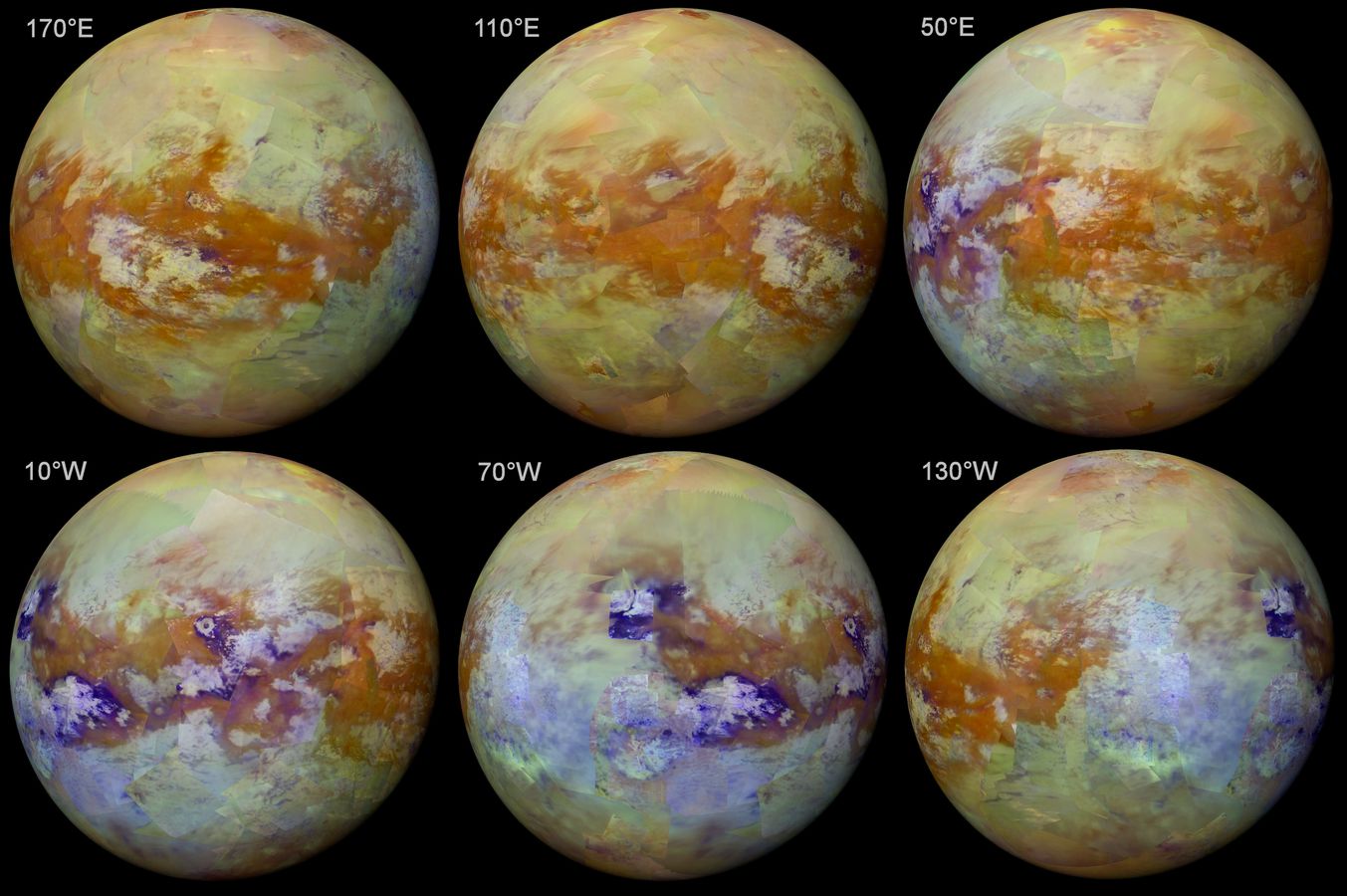}
 \caption{Selection of orthographic views derived from the RGB corrected color ratio map of \figref{fig_9}b after a last cosmetic hand cleaning step. The upper left image is centered at (\ang{0}, \ang{170}E), close to the Huygens landing site. The six panels correspond to views where Titan is rotated by 60 deg. in longitude eastward from left to right and from top to bottom. The equatorial dune fields appear readily in brown. Bright terrains and several patches of bluish areas also show up in specific locations.}
 \label{fig:fig_10}
\end{figure*}

\figref{fig_9}b is so far the most advanced global cartographic product that we have been able to automatically produce. \figref{fig_10} shows a series of orthographic views centered on the equator, with a point of perspective at infinite distance, derived from this map after a last minor cosmetic hand cleaning step. The six panels correspond to hemispheric views where Titan is rotated by \ang{60} in longitude eastward from left to right and from top to bottom. One of the most striking feature is the equatorial dune fields \citep{Radebaugh2008, Rodriguez2014}, which appear readily in brownish tones. The second main spectral type of interest corresponds to dark blue areas such as the ones we see around Sinlap and Menvra craters, or north east of Hotei Regio. This color difference can be spectrally explained by a local enrichment in water ice, decreasing the reflectance at 1.59 and \SI{2.01}{\um} compared to the \SI{1.27}{\um} channel \citep{Rodriguez2006, McCord2008, Brossier2018, Solomonidou2018}.
However, we point out that this interpretation is not unique, as several organic compounds could potentially create the same spectral effect. Indeed, many organics show the downward trend with increasing wavelength like water ice \citep[\emph{e.g.,}][]{Clark2009, Clark2010, Kokaly2017}. NH-bearing compounds show an even stronger downward trend than water ice. Discriminating between these compositional signatures will require a very precise atmospheric removal and analysis of the detailed spectral structure within each window, which is still an ongoing field of research. Our main objective here was rather to show the global distribution of spectral heterogeneities itself. We leave the identification of individual constituents to further dedicated studies, which could rely both on laboratory spectra and detailed radiative transfer modeling, and which fall beyond the scope of this paper. We now give a regional example of the color map in one of the most important spot on Titan: the Huygens landing site.

\subsection{Zoom on the Huygens landing site}
In order to illustrate the accuracy of the final band ratio map, \figref{fig_11} shows a zoom on the Huygens Landing site. This area is the only spectral measurement acquired from the surface and is therefore of particular interest. The first VIMS observation acquired at Ta in October 2004 (cube CM\_1477491859) had a spatial sampling of \SI{14}{km/pixel} and provided the general context \citep{Rodriguez2006}. The best observation of the landing site itself has then been acquired at T47 in November 2008 (cube labeled CM\_1605804042), when VIMS was operating at closest approach. This allowed to acquire an observation in a spot pointing mode, with the whole spacecraft spinning progressively to compensate for the fast drift of the surface. The spatial sampling of the T47 cube ranged between \num{0.75} and \SI{1.4}{km/pixel}. Other late observations provided the intermediate context. They were acquired at T88 in November 2012 (cube CM\_1732874866, between \num{2.1} and \SI{2.7}{km/pixel}) and T85 in July 2012 (cube CM\_1721856031, between \num{3.2} and \SI{5.5}{km/pixel}).

\begin{figure}[!ht]
 \includegraphics[width=\linewidth]{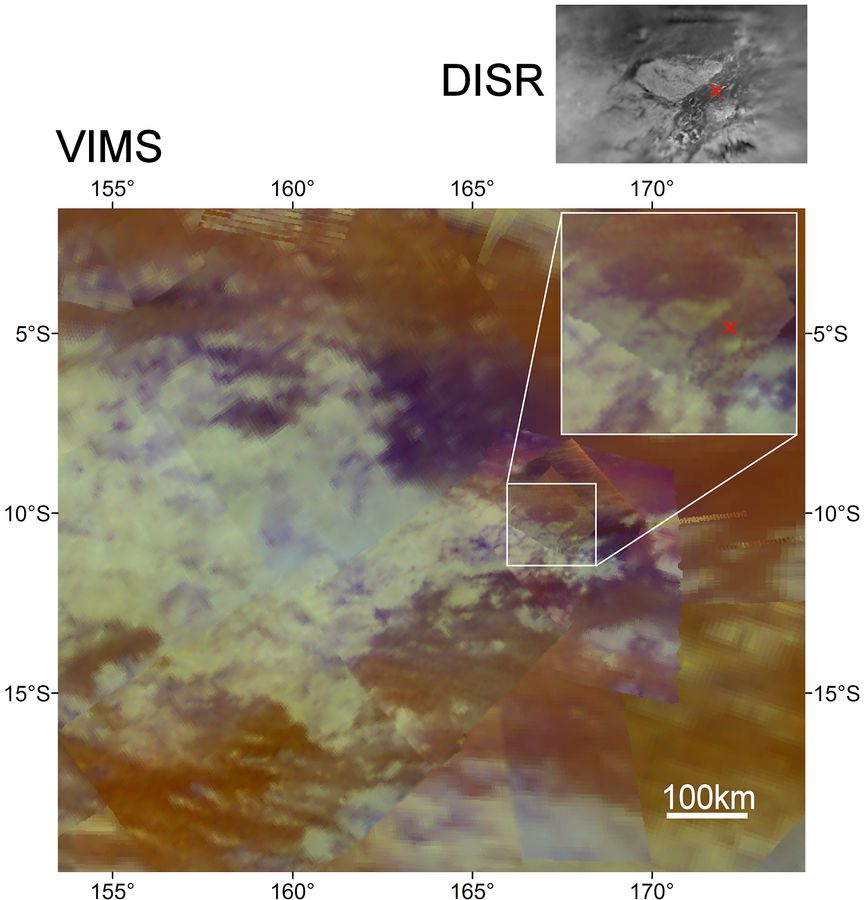}
 \caption{Detail of the VIMS color ratio map corresponding to the area in the white square in \figref{fig_9}b. The Huygens landing site is marked by a red cross. A black and white panorama acquired by DISR on Huygens from an altitude of \SI{\sim 34}{km} is shown for comparison (top). We can easily recognize the bright feature north of the landing site in both data sets. VIMS suggests that Huygens landed in an area corresponding to the moderate bluish tone units.}
 \label{fig:fig_11}
\end{figure}

In \figref{fig_11}, we see that all these data have been homogeneously merged into the color band ratio map, despite being acquired with different geometries and with a time span of eight years. It appears that the Huygens probe landed in an area which corresponds to moderately bluish tones. The accuracy of the VIMS observation is sufficiently high to easily recognize the bright dissected terrain that was imaged by the DISR camera at an altitude of \SI{\sim 34}{km} during the descent of Huygens under its parachute \citep{Tomasko2005, Karkoschka2016}. Further studies could be considered to perform a detailed comparison between the VIMS cubes mentioned here and the DISR data, following the work of \cite{Karkoschka2016}.

\section{Conclusion}

We have reduced the global VIMS hyperspectral archive of Titan integrating data from T0 to T126 flybys in order to map the spectral heterogeneities at the surface. Multispectral summary images have been computed for each hyperspectral VIMS cube, in order to give an easy access to the scientific content of each observation. These browse products are available on the vims.univ-nantes.fr website. Based on automatic filters and manual refinement in the cube selection using these summary images, \num{\sim 19000} individual data cubes have then been merged to produce global color maps at \SI{32}{pixels} per degree (\SI{\sim 1.4}{km/pixel} at the equator) in the seven atmospheric methane optical windows. We implemented a correction for the surface photometric function which takes into account the incidence, emergence, and phase variations. An empirical subtraction of the band wings is used to mitigate the effects of the additive scattering aerosols at short wavelengths. We also investigated band ratios, a powerful technique to emphasize subtle spectral variations, by implementing an empirical correction of the absorption difference between the ratioed channels. This process allowed us to build global maps which integrate data from the complete mission and which strongly emphasize the global distribution of the main spectral units. In particular, the band ratio global map readily shows the extent of the equatorial dune fields (which appears in brown tones in band ratio RGB composites of the \SI{1.59/1.27}{\um}, \SI{2.03/1.27}{\um} and \SI{1.27/1.08}{\um} channels). Several areas show a dark bluish color, such as near Sinlap, Menvra or Selk craters for example, due to a change in composition and/or grain size.

The residual discrepancies in the maps are due to several factors. One of the challenges comes from temporal variations at the surface \citep{Barnes2013b, Solomonidou2016} and moreover in the atmosphere (haze and clouds). This is particularly true at the poles, where significant changes occurred during the mission \citep{LeMouelic2018}. The north pole was fully covered by haze and cloud up to \ang{\sim 55}N at the beginning of the mission. We had to wait for the circulation turnover after the equinox in 2009 to get clearer skies in the north. The south pole experienced a reverse situation, with clear skies at the beginning of the mission and a polar cloud appearing after 2012 and growing in size up to 2017. In addition to these polar events, sporadic methane clouds have been observed throughout all the mission \citep[\emph{e.g.,}][]{Rodriguez2009, Rodriguez2011, Turtle2018}.

The surface photometric behavior can be improved in further studies using a more complex photometric function using wavelength-dependent parameters. This will require a better decorrelation between atmospheric and surface contributions in the methane windows. More inputs derived from a complete radiative transfer analysis could also provide another way to improve the homogeneity of the maps in future works \citep{Cornet2017}.

\section*{Acknowledgment}
Authors are very grateful to two anonymous reviewers for their very detailed comments.
This work has been partly funded by the French spatial agency (CNES).
We also acknowledge the financial support from R\'egion Pays de la Loire, project GeoPlaNet (convention \texttt{2016-10982}).
S.R. is supported by the Institut Universitaire de France and acknowledges support from the UnivEarthS LabEx program of Sorbonne Paris Cit\'e (\texttt{ANR-10-LABX-0023}, \texttt{ANR-11-IDEX-0005-02}) and the French National Research Agency (\texttt{ANR-APOSTIC-11-BS56-002}, \texttt{ANR-12-BS05-001-3/EXO-DUNES}).

\bibliography{Biblio}

\begin{thebibliography}{51}
\providecommand{\natexlab}[1]{#1}
\providecommand{\url}[1]{\texttt{#1}}
\expandafter\ifx\csname urlstyle\endcsname\relax
  \providecommand{\doi}[1]{doi: #1}\else
  \providecommand{\doi}{doi: \begingroup \urlstyle{rm}\Url}\fi

\bibitem[Aharonson et~al.(2014)Aharonson, Hayes, Hayne, Lopes, Lucas, and
  Perron]{Aharonson2014}
Aharonson O. and~5~colleagues.
\newblock
  \href{https://www.cambridge.org/core/product/identifier/CBO9780511667398A013/type/book{\_}part}{{Titan's
  surface geology}}.
\newblock In Muller-Wodarg I. and~3~colleagues, editors, \emph{Titan}, pp
  63--101. Cambridge University Press, Cambridge, {\bf 2014}.

\bibitem[Barnes et~al.(2007)Barnes, Brown, Soderblom, Buratti, Sotin,
  Rodriguez, {Le Mou{\`{e}}lic}, Baines, Clark, and Nicholson]{Barnes2007}
Barnes J.~W. and~9~colleagues.
\newblock \href{http://dx.doi.org/10.1016/j.icarus.2006.08.021}{{Global-scale
  surface spectral variations on Titan seen from Cassini/VIMS}}.
\newblock \emph{Icarus}, 186\penalty0 (1)\penalty0 242--258, {\bf 2007}.

\bibitem[Barnes et~al.(2011)Barnes, Bow, Schwartz, Brown, Soderblom, Hayes,
  Vixie, {Le Mou{\'{e}}lic}, Rodriguez, Sotin, Jaumann, Stephan, Soderblom,
  Clark, Buratti, Baines, and Nicholson]{Barnes2011}
Barnes J.~W. and~16~colleagues.
\newblock \href{http://dx.doi.org/10.1016/j.icarus.2011.08.022}{{Organic
  sedimentary deposits in Titan's dry lakebeds: Probable evaporite}}.
\newblock \emph{Icarus}, 216\penalty0 (1)\penalty0 136--140, {\bf 2011}.

\bibitem[Barnes et~al.(2013{\natexlab{a}})Barnes, Buratti, Turtle, Bow, Dalba,
  Perry, Brown, Rodriguez, Mou{\'{e}}lic, Baines, Sotin, Lorenz, Malaska,
  McCord, Clark, Jaumann, Hayne, Nicholson, Soderblom, and
  Soderblom]{Barnes2013b}
Barnes J.~W. and~19~colleagues.
\newblock \href{http://dx.doi.org/10.1186/2191-2521-2-1}{{Precipitation-induced
  surface brightenings seen on Titan by Cassini VIMS and ISS}}.
\newblock \emph{Planetary Science}, 2\penalty0 (1)\penalty0 1, {\bf
  2013{\natexlab{a}}}.

\bibitem[Barnes et~al.(2013{\natexlab{b}})Barnes, Clark, Sotin,
  {\'{A}}d{\'{a}}mkovics, App{\'{e}}r{\'{e}}, Rodriguez, Soderblom, Brown,
  Buratti, Baines, {Le Mou{\'{e}}lic}, and Nicholson]{Barnes2013a}
Barnes J.~W. and~11~colleagues.
\newblock \href{http://dx.doi.org/10.1088/0004-637X/777/2/161}{{A transmission
  spectrum of titan's north polar atmosphere from a specular reflection of the
  sun}}.
\newblock \emph{Astrophysical Journal}, 777\penalty0 (2), {\bf
  2013{\natexlab{b}}}.

\bibitem[Barnes et~al.(2014)Barnes, Sotin, Soderblom, Brown, Hayes, Donelan,
  Rodriguez, Mou{\'{e}}lic, Baines, and McCord]{Barnes2014}
Barnes J.~W. and~9~colleagues.
\newblock \href{http://dx.doi.org/10.1186/s13535-014-0003-4}{{Cassini/VIMS
  observes rough surfaces on Titan's Punga Mare in specular reflection}}.
\newblock \emph{Planetary Science}, 3\penalty0 (1)\penalty0 3, {\bf 2014}.

\bibitem[Binder and Jones(1972)]{Binder1972}
Binder A.~B. and Jones J.~C.
\newblock \href{http://dx.doi.org/10.1029/JB077i017p03005}{{Spectrophotometric
  studies of the photometric function, composition, and distribution of the
  surface materials of Mars}}.
\newblock \emph{Journal of Geophysical Research}, 77\penalty0 (17)\penalty0
  3005--3020, {\bf 1972}.

\bibitem[Brossier et~al.(2018)Brossier, Rodriguez, Cornet, Lucas, Radebaugh,
  Maltagliati, {Le Mou{\'{e}}lic}, Solomonidou, Coustenis, Hirtzig, Jaumann,
  Stephan, and Sotin]{Brossier2018}
Brossier J.~F. and~12~colleagues.
\newblock \href{http://dx.doi.org/10.1029/2017JE005399}{{Geological Evolution
  of Titan's Equatorial Regions: Possible Nature and Origin of the Dune
  Material}}.
\newblock \emph{Journal of Geophysical Research: Planets}, 123\penalty0
  (5)\penalty0 1089--1112, {\bf 2018}.

\bibitem[Brown et~al.(2010)Brown, Roberts, and Schaller]{Brown2010}
Brown M.~E., Roberts J.~E. and Schaller E.~L.
\newblock \href{http://dx.doi.org/10.1016/j.icarus.2009.08.024}{{Clouds on
  Titan during the Cassini prime mission: A complete analysis of the VIMS
  data}}.
\newblock \emph{Icarus}, 205\penalty0 (2)\penalty0 571--580, {\bf 2010}.

\bibitem[Brown et~al.(2004)Brown, Baines, Bellucci, Bibring, Buratti,
  Capaccioni, Cerroni, Clark, Coradini, Cruikshank, Drossart, Formisano,
  Jaumann, Langevin, Matson, Mccord, Mennella, Miller, Nelson, Nicholson,
  Sicardy, and Sotin]{Brown2004}
Brown R.~H. and~21~colleagues.
\newblock \href{http://link.springer.com/10.1007/1-4020-3874-7{\_}3}{{The
  Cassini Visual and Infrared Mapping Spectrometer (VIMS) Investigation}}.
\newblock In \emph{The Cassini-Huygens Mission}, pp 111--168. Kluwer Academic
  Publishers, Dordrecht, {\bf 2004}.

\bibitem[Clark et~al.(2009)Clark, Curchin, Hoefen, and Swayze]{Clark2009}
Clark R.~N. and~3~colleagues.
\newblock \href{http://dx.doi.org/10.1029/2008JE003150}{{Reflectance
  spectroscopy of organic compounds: 1. Alkanes}}.
\newblock \emph{Journal of Geophysical Research}, 114\penalty0 (E3)\penalty0
  E03001, {\bf 2009}.

\bibitem[Clark et~al.(2010)Clark, Curchin, Barnes, Jaumann, Soderblom,
  Cruikshank, Brown, Rodriguez, Lunine, Stephan, Hoefen, {Le Mou{\'{e}}lic},
  Sotin, Baines, Buratti, and Nicholson]{Clark2010}
Clark R.~N. and~15~colleagues.
\newblock \href{http://dx.doi.org/10.1029/2009JE003369}{{Detection and mapping
  of hydrocarbon deposits on Titan}}.
\newblock \emph{Journal of Geophysical Research}, 115\penalty0 (10), {\bf
  2010}.

\bibitem[Clark et~al.(2018)Clark, Brown, Lytle, and Hedman]{Clark2018}
Clark R.~N. and~3~colleagues.
\newblock
  \href{http://atmos.nmsu.edu/data{\_}and{\_}services/atmospheres{\_}data/Cassini/vims.html}{{The
  VIMS Wavelength and Radiometric Calibration 19, Final Report}}.
\newblock The Planetary Atmospheres Node, {\bf 2018}.

\bibitem[Cornet et~al.(2012)Cornet, Bourgeois, {Le Mou{\'{e}}lic}, Rodriguez,
  {Lopez Gonzalez}, Sotin, Tobie, Fleurant, Barnes, Brown, Baines, Buratti,
  Clark, and Nicholson]{Cornet2012}
Cornet T. and~13~colleagues.
\newblock
  \href{http://dx.doi.org/10.1016/j.icarus.2012.01.013}{{Geomorphological
  significance of Ontario Lacus on Titan: Integrated interpretation of Cassini
  VIMS, ISS and RADAR data and comparison with the Etosha Pan (Namibia)}}.
\newblock \emph{Icarus}, 218\penalty0 (2)\penalty0 788--806, {\bf 2012}.

\bibitem[Cornet et~al.(2017)Cornet, Rodriguez, Maltagliati, App{\'{e}}r{\'{e}},
  Sotin, {Le Mou{\'{e}}lic}, Rannou, Solomonidou, Hirtzig, B{\'{e}}zard,
  Coustenis, Brown, Barnes, Baines, Buratti, Clark, and Nicholson]{Cornet2017}
Cornet T. and~16~colleagues.
\newblock \href{http://adsabs.harvard.edu/abs/2017LPI....48.1847C}{{Radiative
  Transfer Modelling in Titan's Atmosphere: Application to Cassini/VIMS Data}}.
\newblock In \emph{48th Lunar and Planetary Science Conference}, volume~48,
  Texas, {\bf 2017}.

\bibitem[Elachi et~al.(2005)Elachi, Wall, Allison, Anderson, Boehmer, Callahan,
  Encrenaz, Flamini, Franceschetti, Gim, Hamilton, Hensley, Janssen, Johnson,
  Kelleher, Kirk, Lopes, Lorenz, Lunine, Muhleman, Ostro, Paganelli, Picardi,
  Posa, Roth, Seu, Shaffer, Soderblom, Stiles, Stofan, Vetrella, West, Wood,
  Wye, and Zebker]{Elachi2005}
Elachi C. and~34~colleagues.
\newblock \href{http://dx.doi.org/10.1126/science.1109919}{{Cassini Radar Views
  the Surface of Titan}}.
\newblock \emph{Science}, 308\penalty0 (5724)\penalty0 970--974, {\bf 2005}.

\bibitem[Hapke(2012)]{Hapke2012}
Hapke B.
\newblock
  \href{http://ebooks.cambridge.org/ref/id/CBO9781139025683}{\emph{{Theory of
  Reflectance and Emittance Spectroscopy}}}.
\newblock Cambridge University Press, Cambridge, 2nd edition, {\bf 2012}.

\bibitem[Hapke(1963)]{Hapke1963}
Hapke B.~W.
\newblock \href{http://dx.doi.org/10.1029/JZ068i015p04571}{{A theoretical
  photometric function for the lunar surface}}.
\newblock \emph{Journal of Geophysical Research}, 68\penalty0 (15)\penalty0
  4571--4586, {\bf 1963}.

\bibitem[Hirtzig et~al.(2009)Hirtzig, Tokano, Rodriguez, le~Mou{\'{e}}lic, and
  Sotin]{Hirtzig2009}
Hirtzig M. and~4~colleagues.
\newblock \href{http://dx.doi.org/10.1007/s00159-009-0018-0}{{A review of
  Titan's atmospheric phenomena}}.
\newblock \emph{The Astronomy and Astrophysics Review}, 17\penalty0
  (2)\penalty0 105--147, {\bf 2009}.

\bibitem[Karkoschka and Schr{\"{o}}der(2016)]{Karkoschka2016}
Karkoschka E. and Schr{\"{o}}der S.~E.
\newblock \href{http://dx.doi.org/10.1016/j.icarus.2015.06.010}{{Eight-color
  maps of Titan's surface from spectroscopy with Huygens' DISR}}.
\newblock \emph{Icarus}, 270\penalty0 260--271, {\bf 2016}.

\bibitem[Kokaly et~al.(2017)Kokaly, Clark, Swayze, Livo, Hoefen, Pearson, Wise,
  Benzel, Lowers, Driscoll, and Klein]{Kokaly2017}
Kokaly R.~F. and~10~colleagues.
\newblock \href{http://dx.doi.org/10.3133/ds1035}{{USGS Spectral Library
  Version 7}}.
\newblock \emph{Data Series}, p~61, {\bf 2017}.

\bibitem[Lane and Irvine(1973)]{Lane1973}
Lane A.~P. and Irvine W.~M.
\newblock \href{http://dx.doi.org/10.1086/111414}{{Monochromatic phase curves
  and albedos for the lunar disk}}.
\newblock \emph{The Astronomical Journal}, 78\penalty0 (1962)\penalty0 267,
  {\bf 1973}.

\bibitem[{Le Mou{\'{e}}lic} et~al.(2018){Le Mou{\'{e}}lic}, Rodriguez, Robidel,
  Rousseau, Seignovert, Sotin, Barnes, Brown, Baines, Buratti, Clark,
  Nicholson, Rannou, and Cornet]{LeMouelic2018}
{Le Mou{\'{e}}lic} S. and~13~colleagues.
\newblock \href{http://dx.doi.org/10.1016/j.icarus.2018.04.028}{{Mapping polar
  atmospheric features on Titan with VIMS: From the dissipation of the northern
  cloud to the onset of a southern polar vortex}}.
\newblock \emph{Icarus}, 311\penalty0 371--383, {\bf 2018}.

\bibitem[{Le Mou{\'{e}}lic} et~al.(2008){Le Mou{\'{e}}lic}, Paillou, Janssen,
  Barnes, Rodriguez, Sotin, Brown, Baines, Buratti, Clark, Crapeau, Encrenaz,
  Jaumann, Geudtner, Paganelli, Soderblom, Tobie, and Wall]{LeMouelic2008}
{Le Mou{\'{e}}lic} S. and~17~colleagues.
\newblock \href{http://dx.doi.org/10.1029/2007JE002965}{{Mapping and
  interpretation of Sinlap crater on Titan using Cassini VIMS and RADAR data}}.
\newblock \emph{Journal of Geophysical Research}, 113\penalty0 (E4)\penalty0
  E04003, {\bf 2008}.

\bibitem[{Le Mou{\'{e}}lic} et~al.(2012{\natexlab{a}}){Le Mou{\'{e}}lic},
  Cornet, Rodriguez, Sotin, Barnes, Baines, Brown, Lef{\`{e}}vre, Buratti,
  Clark, and Nicholson]{LeMouelic2012a}
{Le Mou{\'{e}}lic} S. and~10~colleagues.
\newblock \href{http://dx.doi.org/10.1016/j.pss.2012.09.008}{{Global mapping of
  Titan's surface using an empirical processing method for the atmospheric and
  photometric correction of Cassini/VIMS images}}.
\newblock \emph{Planetary and Space Science}, 73\penalty0 (1)\penalty0
  178--190, {\bf 2012{\natexlab{a}}}.

\bibitem[{Le Mou{\'{e}}lic} et~al.(2012{\natexlab{b}}){Le Mou{\'{e}}lic},
  Rannou, Rodriguez, Sotin, Griffith, {Le Corre}, Barnes, Brown, Baines,
  Buratti, Clark, Nicholson, and Tobie]{LeMouelic2012b}
{Le Mou{\'{e}}lic} S. and~12~colleagues.
\newblock \href{http://dx.doi.org/10.1016/j.pss.2011.04.006}{{Dissipation of
  Titans north polar cloud at northern spring equinox}}.
\newblock \emph{Planetary and Space Science}, 60\penalty0 (1)\penalty0 86--92,
  {\bf 2012{\natexlab{b}}}.

\bibitem[Lopes et~al.(2013)Lopes, Kirk, Mitchell, LeGall, Barnes, Hayes,
  Kargel, Wye, Radebaugh, Stofan, Janssen, Neish, Wall, Wood, Lunine, and
  Malaska]{Lopes2013}
Lopes R.~M.~C. and~15~colleagues.
\newblock \href{http://dx.doi.org/10.1002/jgre.20062}{{Cryovolcanism on Titan:
  New results from Cassini RADAR and VIMS}}.
\newblock \emph{Journal of Geophysical Research: Planets}, 118\penalty0
  (3)\penalty0 416--435, {\bf 2013}.

\bibitem[MacKenzie et~al.(2014)MacKenzie, Barnes, Sotin, Soderblom, {Le
  Mou{\'{e}}lic}, Rodriguez, Baines, Buratti, Clark, Nicholson, and
  McCord]{MacKenzie2014}
MacKenzie S.~M. and~10~colleagues.
\newblock \href{http://dx.doi.org/10.1016/j.icarus.2014.08.022}{{Evidence of
  Titan's climate history from evaporite distribution}}.
\newblock \emph{Icarus}, 243\penalty0 191--207, {\bf 2014}.

\bibitem[McCord et~al.(2008)McCord, Hayne, Combe, Hansen, Barnes, Rodriguez,
  {Le Mou{\'{e}}lic}, Baines, Buratti, Sotin, Nicholson, Jaumann, Nelson, and
  {the Cassini VIMS Team}]{McCord2008}
McCord T.~B. and~13~colleagues.
\newblock \href{http://dx.doi.org/10.1016/j.icarus.2007.08.039}{{Titan's
  surface: Search for spectral diversity and composition using the Cassini VIMS
  investigation}}.
\newblock \emph{Icarus}, 194\penalty0 (1)\penalty0 212--242, {\bf 2008}.

\bibitem[Neish et~al.(2016)Neish, Molaro, Lora, Howard, Kirk, Schenk, Bray, and
  Lorenz]{Neish2016}
Neish C.~D. and~7~colleagues.
\newblock \href{http://dx.doi.org/10.1016/j.icarus.2015.07.022}{{Fluvial
  erosion as a mechanism for crater modification on Titan}}.
\newblock \emph{Icarus}, 270\penalty0 114--129, {\bf 2016}.

\bibitem[Pelkey et~al.(2007)Pelkey, Mustard, Murchie, Clancy, Wolff, Smith,
  Milliken, Bibring, Gendrin, Poulet, Langevin, and Gondet]{Pelkey2007}
Pelkey S.~M. and~11~colleagues.
\newblock \href{http://dx.doi.org/10.1029/2006JE002831}{{CRISM multispectral
  summary products: Parameterizing mineral diversity on Mars from
  reflectance}}.
\newblock \emph{Journal of Geophysical Research}, 112\penalty0 (E8)\penalty0
  E08S14, {\bf 2007}.

\bibitem[Porco et~al.(2005)Porco, Baker, Barbara, Beurle, Brahic, Burns,
  Charnoz, Cooper, Dawson, {Del Genio}, Denk, Dones, Dyudina, Evans, Fussner,
  Giese, Grazier, Helfenstein, Ingersoll, Jacobson, Johnson, McEwen, Murray,
  Neukum, Owen, Perry, Roatsch, Spitale, Squyres, Thomas, Tiscareno, Turtle,
  Vasavada, Veverka, Wagner, and West]{Porco2005}
Porco C.~C. and~35~colleagues.
\newblock \href{http://dx.doi.org/10.1038/nature03436}{{Imaging of Titan from
  the Cassini spacecraft.}}
\newblock \emph{Nature}, 434\penalty0 159--168, {\bf 2005}.

\bibitem[Radebaugh et~al.(2007)Radebaugh, Lorenz, Kirk, Lunine, Stofan, Lopes,
  and Wall]{Radebaugh2007}
Radebaugh J. and~6~colleagues.
\newblock \href{http://dx.doi.org/10.1016/j.icarus.2007.06.020}{{Mountains on
  Titan observed by Cassini Radar}}.
\newblock \emph{Icarus}, 192\penalty0 (1)\penalty0 77--91, {\bf 2007}.

\bibitem[Radebaugh et~al.(2008)Radebaugh, Lorenz, Lunine, Wall, Boubin, Reffet,
  Kirk, Lopes, Stofan, Soderblom, Allison, Janssen, Paillou, Callahan, Spencer,
  and {the Cassini Radar Team}]{Radebaugh2008}
Radebaugh J. and~15~colleagues.
\newblock \href{http://dx.doi.org/10.1016/j.icarus.2007.10.015}{{Dunes on Titan
  observed by Cassini Radar}}.
\newblock \emph{Icarus}, 194\penalty0 (2)\penalty0 690--703, {\bf 2008}.

\bibitem[Rodriguez et~al.(2006)Rodriguez, {Le Mou{\'{e}}lic}, Sotin,
  Cl{\'{e}}net, Clark, Buratti, Brown, McCord, Nicholson, and
  Baines]{Rodriguez2006}
Rodriguez S. and~9~colleagues.
\newblock \href{http://dx.doi.org/10.1016/j.pss.2006.06.016}{{Cassini/VIMS
  hyperspectral observations of the HUYGENS landing site on Titan}}.
\newblock \emph{Planetary and Space Science}, 54\penalty0 (15)\penalty0
  1510--1523, {\bf 2006}.

\bibitem[Rodriguez et~al.(2011)Rodriguez, {Le Mou{\'{e}}lic}, Rannou, Sotin,
  Brown, Barnes, Griffith, Burgalat, Baines, Buratti, Clark, and
  Nicholson]{Rodriguez2011}
Rodriguez S. and~11~colleagues.
\newblock \href{http://dx.doi.org/10.1016/j.icarus.2011.07.031}{{Titan's cloud
  seasonal activity from winter to spring with Cassini/VIMS}}.
\newblock \emph{Icarus}, 216\penalty0 (1)\penalty0 89--110, {\bf 2011}.

\bibitem[Rodriguez et~al.(2014)Rodriguez, Garcia, Lucas, App{\'{e}}r{\'{e}},
  {Le Gall}, Reffet, {Le Corre}, {Le Mou{\'{e}}lic}, Cornet, {Courrech du
  Pont}, Narteau, Bourgeois, Radebaugh, Arnold, Barnes, Stephan, Jaumann,
  Sotin, Brown, Lorenz, and Turtle]{Rodriguez2014}
Rodriguez S. and~20~colleagues.
\newblock \href{http://dx.doi.org/10.1016/j.icarus.2013.11.017}{{Global mapping
  and characterization of Titan's dune fields with Cassini: Correlation between
  RADAR and VIMS observations}}.
\newblock \emph{Icarus}, 230\penalty0 168--179, {\bf 2014}.

\bibitem[Rodriguez et~al.(2009)Rodriguez, {Le Mou{\'{e}}lic}, Rannou, Tobie,
  Baines, Barnes, Griffith, Hirtzig, Pitman, Sotin, Brown, Buratti, Clark, and
  Nicholson]{Rodriguez2009}
Rodriguez S. and~13~colleagues.
\newblock \href{http://dx.doi.org/10.1038/nature08014}{{Global circulation as
  the main source of cloud activity on Titan}}.
\newblock \emph{Nature}, 459\penalty0 (7247)\penalty0 678--682, {\bf 2009}.

\bibitem[Seignovert(2015)]{Seignovert2015}
Seignovert B.
\newblock \href{https://doi.org/10.5281/zenodo.155767}{{Cassini Titan flyby}},
  {\bf 2015}.

\bibitem[Soderblom et~al.(2012)Soderblom, Barnes, Soderblom, Brown, Griffith,
  Nicholson, Stephan, Jaumann, Sotin, Baines, Buratti, and
  Clark]{Soderblom2012}
Soderblom J.~M. and~11~colleagues.
\newblock \href{http://dx.doi.org/10.1016/j.icarus.2012.05.030}{{Modeling
  specular reflections from hydrocarbon lakes on Titan}}.
\newblock \emph{Icarus}, 220\penalty0 (2)\penalty0 744--751, {\bf 2012}.

\bibitem[Soderblom et~al.(2009{\natexlab{a}})Soderblom, Barnes, Brown, Clark,
  Janssen, McCord, Niemann, and Tomasko]{Soderblom2009b}
Soderblom L.~A. and~7~colleagues.
\newblock \href{https://doi.org/10.1007/978-1-4020-9215-2{\_}6
  http://link.springer.com/10.1007/978-1-4020-9215-2{\_}6}{\emph{{Composition
  of Titan's Surface}}}, pp 141--175.
\newblock Springer Netherlands, Dordrecht, {\bf 2009{\natexlab{a}}}.

\bibitem[Soderblom et~al.(2009{\natexlab{b}})Soderblom, Brown, Soderblom,
  Barnes, Kirk, Sotin, Jaumann, Mackinnon, Mackowski, Baines, Buratti, Clark,
  and Nicholson]{Soderblom2009a}
Soderblom L.~A. and~12~colleagues.
\newblock \href{http://dx.doi.org/10.1016/j.icarus.2009.07.033}{{The geology of
  Hotei Regio, Titan: Correlation of Cassini VIMS and RADAR}}.
\newblock \emph{Icarus}, 204\penalty0 (2)\penalty0 610--618, {\bf
  2009{\natexlab{b}}}.

\bibitem[Solomonidou et~al.(2018)Solomonidou, Coustenis, Lopes, Malaska,
  Rodriguez, Drossart, Elachi, Schmitt, Philippe, Janssen, Hirtzig, Wall,
  Sotin, Lawrence, Altobelli, Bratsolis, Radebaugh, Stephan, Brown, {Le
  Mou{\'{e}}lic}, {Le Gall}, Villanueva, Brossier, Bloom, Witasse, Matsoukas,
  and Schoenfeld]{Solomonidou2018}
Solomonidou A. and~26~colleagues.
\newblock \href{http://dx.doi.org/10.1002/2017JE005477}{{The Spectral Nature of
  Titan's Major Geomorphological Units: Constraints on Surface Composition}}.
\newblock \emph{Journal of Geophysical Research: Planets}, 123\penalty0
  (2)\penalty0 489--507, {\bf 2018}.

\bibitem[Solomonidou et~al.(2016)Solomonidou, Coustenis, Hirtzig, Rodriguez,
  Stephan, Lopes, Drossart, Sotin, {Le Mou{\'{e}}lic}, Lawrence, Bratsolis,
  Jaumann, and Brown]{Solomonidou2016}
Solomonidou A. and~12~colleagues.
\newblock \href{http://dx.doi.org/10.1016/j.icarus.2015.05.003}{{Temporal
  variations of Titan's surface with Cassini/VIMS}}.
\newblock \emph{Icarus}, 270\penalty0 85--99, {\bf 2016}.

\bibitem[Sotin et~al.(2005)Sotin, Jaumann, Buratti, Brown, Clark, Soderblom,
  Baines, Bellucci, Bibring, Capaccioni, Cerroni, Combes, Coradini, Cruikshank,
  Drossart, Formisano, Langevin, Matson, McCord, Nelson, Nicholson, Sicardy,
  LeMou{\'{e}}lic, Rodriguez, Stephan, and Scholz]{Sotin2005}
Sotin C. and~25~colleagues.
\newblock \href{http://dx.doi.org/10.1038/nature03596}{{Release of volatiles
  from a possible cryovolcano from near-infrared imaging of Titan.}}
\newblock \emph{Nature}, 435\penalty0 (7043)\penalty0 786--789, {\bf 2005}.

\bibitem[Sotin et~al.(2012)Sotin, Lawrence, Reinhardt, Barnes, Brown, Hayes,
  {Le Mou{\'{e}}lic}, Rodriguez, Soderblom, Soderblom, Baines, Buratti, Clark,
  Jaumann, Nicholson, and Stephan]{Sotin2012}
Sotin C. and~15~colleagues.
\newblock \href{http://dx.doi.org/10.1016/j.icarus.2012.08.017}{{Observations
  of Titan's Northern lakes at 5$\mu$m: Implications for the organic cycle and
  geology}}.
\newblock \emph{Icarus}, 221\penalty0 (2)\penalty0 768--786, {\bf 2012}.

\bibitem[Stofan et~al.(2007)Stofan, Elachi, Lunine, Lorenz, Stiles, Mitchell,
  Ostro, Soderblom, Wood, Zebker, Wall, Janssen, Kirk, Lopes, Paganelli,
  Radebaugh, Wye, Anderson, Allison, Boehmer, Callahan, Encrenaz, Flamini,
  Francescetti, Gim, Hamilton, Hensley, Johnson, Kelleher, Muhleman, Paillou,
  Picardi, Posa, Roth, Seu, Shaffer, Vetrella, and West]{Stofan2007}
Stofan E.~R. and~37~colleagues.
\newblock \href{http://dx.doi.org/10.1038/nature05438}{{The lakes of Titan}}.
\newblock \emph{Nature}, 445\penalty0 (7123)\penalty0 61--64, {\bf 2007}.

\bibitem[Tomasko et~al.(2005)Tomasko, Archinal, Becker, B{\'{e}}zard, Bushroe,
  Combes, Cook, Coustenis, de~Bergh, Dafoe, Doose, Dout{\'{e}}, Eibl, Engel,
  Gliem, Grieger, Holso, Howington-Kraus, Karkoschka, Keller, Kirk, Kramm,
  K{\"{u}}ppers, Lanagan, Lellouch, Lemmon, Lunine, McFarlane, Moores, Prout,
  Rizk, Rosiek, Rueffer, Schr{\"{o}}der, Schmitt, See, Smith, Soderblom,
  Thomas, and West]{Tomasko2005}
Tomasko M.~G. and~39~colleagues.
\newblock \href{http://dx.doi.org/10.1038/nature04126}{{Rain, winds and haze
  during the Huygens probe's descent to Titan's surface}}.
\newblock \emph{Nature}, 438\penalty0 (7069)\penalty0 765--778, {\bf 2005}.

\bibitem[Turtle et~al.(2018)Turtle, Perry, Barbara, {Del Genio}, Rodriguez, {Le
  Mou{\'{e}}lic}, Sotin, Lora, Faulk, Corlies, Kelland, MacKenzie, West,
  McEwen, Lunine, Pitesky, Ray, and Roy]{Turtle2018}
Turtle E.~P. and~17~colleagues.
\newblock \href{http://dx.doi.org/10.1029/2018GL078170}{{Titan's Meteorology
  Over the Cassini Mission: Evidence for Extensive Subsurface Methane
  Reservoirs}}.
\newblock \emph{Geophysical Research Letters}, 45\penalty0 (11)\penalty0
  5320--5328, {\bf 2018}.

\bibitem[Vixie et~al.(2012)Vixie, Barnes, Bow, {Le Mou{\'{e}}lic}, Rodriguez,
  Brown, Cerroni, Tosi, Buratti, Sotin, Filacchione, Capaccioni, and
  Coradini]{Vixie2012}
Vixie G. and~12~colleagues.
\newblock \href{http://dx.doi.org/10.1016/j.pss.2011.03.021}{{Mapping Titan's
  surface features within the visible spectrum via Cassini VIMS}}.
\newblock \emph{Planetary and Space Science}, 60\penalty0 (1)\penalty0 52--61,
  {\bf 2012}.

\bibitem[Wood et~al.(2010)Wood, Lorenz, Kirk, Lopes, Mitchell, and
  Stofan]{Wood2010}
Wood C.~A. and~5~colleagues.
\newblock \href{http://dx.doi.org/10.1016/j.icarus.2009.08.021}{{Impact craters
  on Titan}}.
\newblock \emph{Icarus}, 206\penalty0 (1)\penalty0 334--344, {\bf 2010}.

\end{thebibliography}

\end{document}